%% file: main.tex
\documentclass[prd,aps,tightenlines,notitlepage,nofootinbib,preprintnumbers,letterpaper,superscriptaddress]{revtex4-2} 

\pdfoutput=1

\input{Packages}

\input{Commands}

\input{units}
\begin{document}

\begin{CJK*}{UTF8}{gbsn}
\title{Sensitivity of Hyper-Kamiokande to sub-eV Sterile Neutrinos}

\author{Emilse~Cabrera\orcidC{}}
\email{emilsecc@aluno.puc-rio.br}
\affiliation{Departamento de F\'isica, Pontif\'icia Universidade Cat\'olica do Rio de Janeiro,\\
Rua Marquês de São Vicente 225, Rio de Janeiro, Brazil}

\author{Arman~Esmaili\orcidB{}}
\email{arman@puc-rio.br}
\affiliation{Departamento de F\'isica, Pontif\'icia Universidade Cat\'olica do Rio de Janeiro,\\
Rua Marquês de São Vicente 225, Rio de Janeiro, Brazil}

\author{Hiroshi~Nunokawa\orcidD{}}
\email{nunokawa@puc-rio.br}
\affiliation{Departamento de F\'isica, Pontif\'icia Universidade Cat\'olica do Rio de Janeiro,\\
Rua Marquês de São Vicente 225, Rio de Janeiro, Brazil}

\author{Ana Maria Garcia~Trzeciak\orcidA{}}
\email{anatrzeciak@puc-rio.br}
\affiliation{Departamento de F\'isica, Pontif\'icia Universidade Cat\'olica do Rio de Janeiro,\\
Rua Marquês de São Vicente 225, Rio de Janeiro, Brazil}


\begin{abstract}

In this work, we investigate the sensitivity of Hyper-Kamiokande (Hyper-K) to light sterile neutrinos within the $(3+1)$ framework, consisting of three active and one sterile neutrino state. We focus on the regime where the new mass-squared splitting satisfies $\Delta m_{41}^2 \lesssim 1$~eV$^2$, a parameter space complementary to short-baseline sterile-neutrino searches. Using both accelerator and atmospheric neutrino samples, we evaluate the expected capability of Hyper-K to constrain active–sterile mixing. Our results show that Hyper-K can significantly improve current bounds on sterile-neutrino parameters and achieve sensitivity that is  competitive with that of future dedicated experiments. 
    
\end{abstract}

\maketitle
\end{CJK*}

\section{Introduction}

Neutrino oscillation experiments have reported robust results that are virtually consistent with the standard three-active neutrino paradigm, characterized by two independent mass-squared splittings, three mixing angles, and one CP violation phase. This framework successfully explains a wide range of experimental results from solar~\cite{Cleveland:1998nv,SAGE:1999nng,GALLEX:1998kcz,SNO:2002tuh,Borexino:2008dzn,Super-Kamiokande:2023jbt}, atmospheric~\cite{Super-Kamiokande:2017yvm,IceCubeCollaboration:2024ssx}, reactor~\cite{KamLAND:2013rgu,RENO:2018dro,DoubleChooz:2019qbj,DayaBay:2022orm}, and accelerator neutrinos~\cite{
T2K:2021xwb,NOvA:2021nfi}. In addition, measurements from the LEP experiment, based on the $Z$ boson decay width, restrict the number of light-active neutrino species that participate in weak interactions to three~\cite{ALEPH:2005ab}. Despite this success, the possible presence of the so-called sterile neutrinos, which are hypothetical states that do not have Standard Model interactions, remains theoretically well motivated and experimentally viable~\cite{Kopp:2013vaa, Giunti:2019aiy,Acero:2022wqg}.

Experimental hints for sterile neutrinos have emerged from some anomalies related to several short-baseline oscillation experiments. LSND and MiniBooNE collaborations reported an excess of events in the $\nu_{\mu}\rightarrow\nu_e$ appearance channel, which might be consistent with oscillations driven by an eV-scale mass splitting~\cite{LSND:2001aii, MiniBooNE:2013uba, MiniBooNE:2020pnu}. Reactor neutrino experiments have also measured lower fluxes of $\bar{\nu}_e$ than predicted, a discrepancy known as the reactor antineutrino anomaly~\cite{Mention:2011rk, Huber:2011wv}, although more recent re-evaluations of reactor neutrino fluxes indicate some reduction of the discrepancy~\cite{Estienne:2019ujo,Kopeikin:2021ugh}. Similarly, experiments with $^{51}$Cr and $^{37}$Ar sources, such as GALLEX and SAGE, observed a deficit in the detection rates of $\nu_e$~\cite{SAGE:1998fvr, Abdurashitov:2005tb}. All of these anomalies can be interpreted as possible signatures of oscillations involving sterile neutrinos with $\Delta m_{41}^2=m_4^2-m_1^2 \sim 1\;\rm{eV}^2$, where $m_i$ are the neutrino masses with $m_4$ being the mostly sterile state. 

Besides the eV-scale mass splitting, there is growing interest in the possibility of very-light sterile neutrinos, with mass-squared splittings comparable to or smaller than the atmospheric mass-squared splitting. The early results of solar neutrino experiments, such as Borexino~\cite{Borexino:2008fkj}, SNO~\cite{SNO:2009uok}, and Super-Kamiokande (Super-K)~\cite{Super-Kamiokande:2010tar}, hinted at the absence of expected upturn at low energies in the solar neutrino spectrum, which potentially can be resolved by the existence of sub-eV sterile neutrinos~\cite{deHolanda:2003tx,deHolanda:2010am}. In this regime, with $\Delta m_{41}^2 \sim 10^{-3}\;\rm{eV}^2$, sterile neutrinos can affect oscillation patterns in long-baseline and atmospheric neutrino experiments~\cite{Kelly_2017,KM3NeT:2021uez,Thakore:2018lgn}. In particular, small splittings can lead to matter-enhanced features in the oscillation probabilities for Earth-scale distances and GeV energies. Atmospheric neutrinos are ideal for probing these small mass splittings.

Sterile neutrino searches have been carried out across a wide range of mass-squared splittings, using all available neutrino sources and baselines~\cite{Acero:2022wqg}. At long baselines, T2K experiment has investigated $\nu_e$ disappearance at its near detector and searched for active-sterile oscillations at Super-K via both charged- and neutral-current interactions~\cite{Himmel:2015cna, T2K:2014xvp, T2K:2019efw}. These analyses, based on beam and atmospheric neutrinos, have primarily targeted the high-$\Delta m_{41}^2$ regime, where active-sterile oscillations can produce measurable distortions. The upcoming Hyper-Kamiokande (Hyper-K) experiment in Japan will significantly improve the sensitivity. Based on the Super-K and T2K detector technology, Hyper-K will feature a larger fiducial volume by nearly an order of magnitude, enhanced detector performance, and an upgraded J-PARC neutrino beam~\cite{Hyper-Kamiokande:2018ofw}. These improvements will allow the detection of subtle oscillation signatures from active-sterile mixing, particularly in the very-light regime with $\Delta m_{41}^2 \lesssim 10^{-3}\;\rm{eV}^2$. In this work, we investigate the Hyper-K sensitivity to very light sterile neutrinos within the $(3+1)$ scenario, with $10^{-5} \lesssim\Delta m_{41}^2/\rm{eV}^2 \lesssim 1$, considering both the beam and atmospheric neutrinos.

The paper is organized as follows. In Section~\ref{sec:pheno}, we present the phenomenology of $(3+1)$ scenario. In Section~\ref{sec:beam}, we describe the simulation of accelerator-based neutrinos at Hyper-K. Section~\ref{sec:atm} is devoted to the atmospheric neutrino analysis, which benefits from a wide energy and baseline coverage. The results, {\it i.e.} the sensitivity of Hyper-K to active–sterile mixing parameters, along with the comparison to existing constraints, are presented in Section~\ref{sec:results}, which also includes our conclusions.

\section{Relevant Phenomenology and Background}\label{sec:pheno}

The $(3+1)$ scenario contains four neutrino states, three standard active neutrinos plus a single sterile neutrino state, where the sterile state does not participate in the weak interactions. The neutrino propagation in flavor basis is governed by the Schr\"odinger-like equation
\begin{equation}
i \frac{{\rm d}}{{\rm d}r} 
\left[
\begin{array}{c}
\nu_e \\
\nu_\mu \\
\nu_\tau \\
\nu_s 
\end{array}
\right]
= 
H(r)
\left[
\begin{array}{c}
\nu_e \\
\nu_\mu \\
\nu_\tau \\
\nu_s 
\end{array}
\right]~,
\label{eq:EvolutionNeutrino}
\end{equation}
where where $r$ indicates the position of neutrino and the Hamiltonian is given by
\begin{equation}
H(r) = \frac{1}{2E_\nu}~ U
\begin{pmatrix}
0 & 0 & 0 & 0 \\
0 & \Delta m_{21}^{2} & 0 & 0 \\
0 & 0 & \Delta m_{31}^{2} & 0 \\
0 & 0 & 0 & \Delta m_{41}^{2}
\end{pmatrix}
U^\dagger
+
\begin{pmatrix}
V_{\rm CC}(r) & 0 & 0 & 0 \\
0 & 0 & 0 & 0 \\
0 & 0 & 0 & 0 \\
0 & 0 & 0 & V_{\rm NC}(r)
\end{pmatrix}~,
\end{equation}
where $U$ is the $4 \times 4$ unitary mixing matrix of the leptonic sector, $\Delta m_{ij}^{2} \equiv m_{i}^{2} - m_{j}^{2}$, and $E_\nu$ is the neutrino energy. The interaction of active neutrinos with matter is represented by the charged current and neutral current potentials, given by
\begin{eqnarray}
V_{\rm CC}(r) &=& \sqrt{2}G_{F}N_{e}(r) \simeq 7.6 \times 10^{-14}~ Y_e \left(\frac{\rho}{\mathrm{g}\hspace{1mm}\mathrm{cm}^{-3}}\right)~\mathrm{eV}~, \\
V_{\rm NC}(r) &=& \sqrt{2}G_{F}N_{n}(r) = \frac{1}{2}\left(\frac{N_{n}(r)}{N_{e}(r)}\right) V_{\rm CC}\simeq \frac{1}{2} V_{\rm CC}~,
\end{eqnarray}
where $G_{F}$ is the Fermi constant, $Y_{e}$ is the number of electrons per nucleon in matter and $N_{e}(r)$ and $N_{n}(r)$ are, respectively, the electron and neutron number density profiles along the neutrino's path of propagation, which are proportional to the matter density denoted by $\rho$. The solution of Eq.~(\ref{eq:EvolutionNeutrino}) is determined by
\begin{equation}
  S = {\mathcal{T}}\mathrm{exp}\left(\int_{0}^{r} H(r')~ {\rm d}r' \right)~,
\end{equation}
where $\mathcal{T}$ indicates the time-ordering operator. The probability of flavor transition $\nu_\alpha \to \nu_\beta$, after traveling distance $r$, is given by
\begin{equation}
 P_{\alpha \beta} = \left|(USU^{\dagger})_{\beta \alpha}\right|^{2}~.
 \label{eq:probEquation}
\end{equation}

For antineutrino oscillation, it is sufficient to replace $U \rightarrow U^{*}$, $V_{\rm CC} \rightarrow -V_{\rm CC}$ and $V_{\rm NC} \rightarrow -V_{\rm NC}$. In the analysis of accelerator neutrinos, the Hamiltonian is assumed to be constant over the entire propagation distance with $\rho = 2.8$ g/cm$^{3}$, and therefore $S = e^{-iHr}$. For propagation through the Earth in the atmospheric neutrino analysis, we use the Preliminary Reference Earth Model (PREM)~\cite{PREM} for density profile and solve Eq.~\eqref{eq:EvolutionNeutrino} by the \texttt{nuSQuIDS} package~\cite{nusquids}.  

We use the following parameterization for the mixing matrix 
\begin{eqnarray}
 U = R(\theta_{34})R(\theta_{24},\delta_{24})R(\theta_{14},\delta_{14})R(\theta_{23})R(\theta_{13},\delta_{13})R(\theta_{12})~,
\end{eqnarray}
with $R$ being the $4 \times 4$ rotation matrices in the $ij$-plane by the rotation angle $\theta_{ij}$, which includes the phase $\delta_{ij}$ for non-adjacent indices. The $4\times4$ mixing matrix $U$ is parameterized by six independent rotation angles $(\theta_{12},\theta_{13},\theta_{23},\theta_{14},\theta_{24},\theta_{34})$, and three independent CP-violating phases $(\delta_{13},\delta_{14},\delta_{24})$, where $\delta_{13}$ corresponds to the unique CP-violating phase in the $3\nu$ scheme.

\subsection{Accelerator Neutrinos\label{sec:pheno_acc}}

\begin{figure}[ht]
    \centering
    \includegraphics[scale = 0.95]{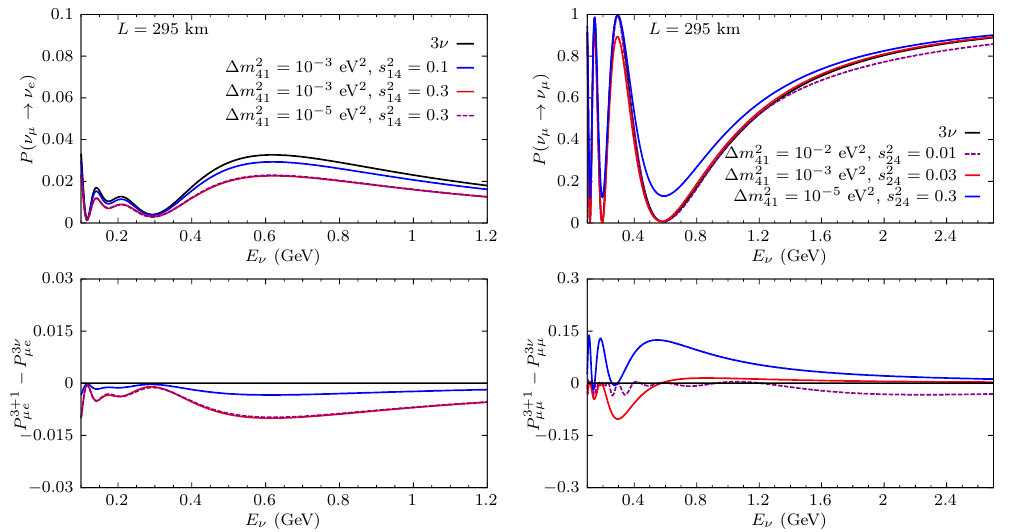}
    \caption{\textit{Top panels:} The electron neutrino appearance (left) and the muon neutrino (right) survival probabilities as function of neutrino energy, setting $L$ = 295~km. \textit{Bottom panels:} differences between the probabilities in $(3+1)$ and $3\nu$ scenarios, as function of neutrino energy. The black curves represents the standard $3\nu$ scheme, while the colored curves correspond to the $(3+1)$ scenario. The chosen values for the standard parameters are shown in Table~\ref{tab:std_osc}. The sterile parameters not displayed in the figure are set to zero. Normal ordering among active neutrinos was assumed. We note that all the curves were obtained by numerical computations without using approximated probability formulas.
    }
    \label{fig:probBeam}
\end{figure}

Inclusion of the fourth sterile state can drastically change the flavor oscillation of active neutrinos over the propagation distance $L=295$~km for accelerator neutrinos at Hyper-K. In our numerical analysis of accelerator neutrinos we precisely compute the oscillation probabilities by taking into account the matter effects. In this section, however, for the purpose of understanding the oscillation patterns, we develop approximate formulae that capture the main features and interpret the oscillation probabilities shown in Figure~\ref{fig:probBeam}. The following simplifying assumptions have been made just for the sake of discussion in this section: -- we assume vacuum propagation for accelerator neutrinos (no matter effect), justified by the subdominant matter effects for these neutrinos for the Hyper-K experiment; -- since for accelerator neutrinos $\Delta m_{21}^2L/4E_\nu\ll1$, for $E_\nu\sim1$~GeV, we assume $\Delta m_{21}^2=0$ and ignore any other mass-squared difference $\sim\mathcal{O}(\Delta m_{21}^2)$. As a consequence, we assume $\Delta m_{31}^2=\Delta m_{32}^2$. Furthermore, here and also in our numerical analysis, throughout this paper, we assume the normal mass ordering ($m_1 < m_2 < m_3)$ among active neutrino mass eigenvalues and set $\delta_{ij} =0$. Also, although the oscillation probabilities in the $\nu_e-\nu_\mu$ system can implicitly depend on $\theta_{34}$ via matter effect, to make the computation time manageable, we assume $\theta_{34}=0$ and focus on the $\theta_{14}$ and $\theta_{24}$ mixing angles. 

With these simplifying assumptions in effect, the oscillation formulae are still cumbersome and difficult to carry on. To capture the essence of oscillation patterns, we consider three representative values of $m_4$, covering the range of interest in our study $10^{-5}\leq\Delta m_{41}^2/{\rm eV}^2 \leq 1$, corresponding to $m_4\sim\mathcal{O}(1)~{\rm eV}$, $m_4\approx m_3$, and $m_4\approx m_2\approx m_1$. We also provide the formulae for a single non-vanishing mixing parameter; {\it i.e.} for $(\theta_{14}\neq0,\theta_{24}=0)$ and $(\theta_{14}=0,\theta_{24}\neq0)$ cases.  

The three relevant oscillation channels in Hyper-K's accelerator analysis are the $\nu_e$-appearance $P(\nu_{\mu}\rightarrow \nu_{e})$, and $\nu_e$- and $\nu_\mu$-survival, $P(\nu_{e}\rightarrow \nu_{e})$ and $P(\nu_{\mu}\rightarrow \nu_{\mu})$, channels. The $\nu_e$-appearance probability, independent of the value of $m_4$, is given by
\begin{eqnarray}
P(\nu_{\mu}\rightarrow \nu_{e}) \approx \cos^{2}\theta_{14}\sin^{2}(2\theta_{{13}})\sin^{2}\theta_{{23}}\sin^{2}\left(\frac{\Delta m^{2}_{31} L}{4E_\nu}\right)~,
\label{eq:Pmue_th24_0}
\end{eqnarray}
for ($\theta_{14} \neq 0, \theta_{24} = 0$), and
\begin{eqnarray}
P(\nu_{\mu}\rightarrow \nu_{e}) \approx \cos^{2}\theta_{24}\sin^{2}(2\theta_{{13}})\sin^{2}\theta_{{23}}  \sin^{2}\left(\frac{\Delta m^{2}_{31} L}{4E_\nu}\right)~,  
\label{eq:Pmue_th14_0}
\end{eqnarray}
for ($\theta_{14} = 0, \theta_{24} \neq 0$). As a consequence of these equations, the $\nu_e$-appearance channel can provide $\Delta m_{41}^2$-independent limits on $\theta_{14}$ and $\theta_{24}$, although the sensitivity is suppressed due to the presence of $\sin^{2}2\theta_{13}$.

In the following, we present the survival oscillation probability formulae for the three representative cases of the value of $m_4$. For the choices ($\theta_{14} \neq 0, \theta_{24} = 0$) and ($\theta_{14} = 0, \theta_{24} \neq 0$), only the $\nu_e$ and $\nu_\mu$-survival probabilities, respectively, are sensitive to sterile-active mixing angles.  \\   

\paragraph{Case I: $m_{4} \sim \mathcal{O}(1)$~eV.} In this case, the mostly-sterile neutrino mass eigenstate has a much larger mass than the three active ones, and so we assume $\Delta m_{41}^{2} \approx \Delta m_{42}^{2} \approx \Delta m_{43}^{2} \equiv \Delta m_{41}^{2}$. For ($\theta_{14} \neq 0, \theta_{24} = 0$), the $\nu_{e}$-survival probability is given by 
\begin{equation}
P(\nu_{e}\rightarrow \nu_{e}) = 1 - \cos^{4}\theta_{14}\sin^{2}(2\theta_{13}) \sin^{2}\left(\frac{\Delta m^{2}_{31} L}{4E_\nu}\right)
- \sin^{2}(2\theta_{14})\sin^{2}\left(\frac{\Delta m^{2}_{41} L}{4E_\nu}\right)~,
\label{eq:Pee_I_th24_0}
\end{equation}
and for ($\theta_{14}=0, \theta_{24} \neq 0$), the $\nu_\mu$-survival probability is
\begin{align}
P(\nu_{\mu}\rightarrow \nu_{\mu}) = 1 &- 4\cos^{4}\theta_{24}\cos^{2}\theta_{13}\sin^{2}\theta_{23}\left[1 - \cos^{2}\theta_{13}\sin^{2}\theta_{23} \right]\sin^{2}\left(\frac{\Delta m^{2}_{31} L}{4E_\nu}\right) \nonumber \\
&- \sin^{2}(2\theta_{24})\sin^{2}\left(\frac{\Delta m^{2}_{41} L}{4E_\nu}\right).
\label{eq:Pmumu_I_th14_0}
\end{align}
The sensitivity to $\theta_{14}$ at $\Delta m_{41}^2\sim1~{\rm eV}^2$ originates from the last term of $\nu_e$-survival probability in Eq.~(\ref{eq:Pee_I_th24_0}), which is not suppressed as the second term by $\sin^2\theta_{13}$ factor. For $\theta_{24}$, the sensitivity comes from both the second and third terms in Eq.~(\ref{eq:Pmumu_I_th14_0}). Notice that for large $m_4$ values, $\sin^{2}(\Delta m^{2}_{41}L/4E_\nu)$ can be replaced by $1/2$, so the effect of $\theta_{14}$ is independent of $\Delta m^{2}_{41}$ and $E_\nu$ values, while $\theta_{24}$ can induce mild energy-dependent features.  \\ 

\paragraph{Case II: $m_{4} \sim m_{3}$.} In this case, at the first approximation, we can set $\Delta m^{2}_{41} \approx \Delta m^{2}_{42} \equiv \Delta m_{41}^{2}$. For ($\theta_{14} \neq 0, \theta_{24} = 0$), we obtain
\begin{align}
P(\nu_{e}\rightarrow \nu_{e}) = 1 &- \cos^{4}\theta_{14}\sin^{2}(2\theta_{13}) \sin^{2}\left(\frac{\Delta m^{2}_{31} L}{4E_\nu}\right) \nonumber \\
&- \sin^{2}(2\theta_{14})\cos^{2}\theta_{13}\sin^{2}\left(\frac{\Delta m^{2}_{41} L}{4E_\nu}\right) \nonumber \\
&- \sin^{2}(2\theta_{14})\sin^{2}\theta_{13}\sin^{2}\left(\frac{\Delta m^{2}_{43} L}{4E_\nu}\right)~,
\label{eq:Pee_II_th24_0}
\end{align}
and for ($\theta_{14}=0, \theta_{24} \neq 0$), we have
\begin{align}
P(\nu_{\mu}\rightarrow \nu_{\mu}) \approx 1 &- 4\cos^{4}\theta_{24}\cos^{2}\theta_{13}\sin^{2}\theta_{23}(1 - \cos^{2}\theta_{13}\sin^{2}\theta_{23})
 \sin^{2}\left(\frac{\Delta m^{2}_{31} L}{4E_\nu}\right) \nonumber \\
&- \sin^{2}(2\theta_{24})\left[1 - \cos^{2}\theta_{13}\sin^{2}\theta_{23}\right]\sin^{2}\left(\frac{\Delta m^{2}_{41} L}{4E_\nu}\right) \nonumber \\
&-\sin^{2}(2\theta_{24})\cos^{2}\theta_{13}\sin^{2}\theta_{23}\sin^{2}\left(\frac{\Delta m^{2}_{43} L}{4E_\nu}\right).
\label{eq:Pmumu_II_th14_0}
\end{align}

In both Eqs.~(\ref{eq:Pee_II_th24_0}) and (\ref{eq:Pmumu_II_th14_0}), the second term with $\sin^2(\Delta m_{31}^2L/4E_\nu)\approx 0.63$ for $E_\nu=0.6$~GeV, provides sensitivity to $\theta_{14}$ and $\theta_{24}$, respectively, although the sensitivity to $\theta_{14}$ is suppressed by the $\sin^2(2\theta_{13})$ factor. The effect of active-sterile mixing appears in the other terms of Eqs.~(\ref{eq:Pee_II_th24_0}) and (\ref{eq:Pmumu_II_th14_0}) when $\Delta m_{(41),(43)}^2 \approx 2\pi E_\nu/L \approx 1.27\times10^{-2}~{\rm eV}^2$, for $E_\nu=0.6$~GeV. Therefore, accelerator neutrinos in Hyper-K can provide competitive sensitivity in the range $10^{-3} < \Delta m_{41}^2/{\rm eV}^2 < 10^{-2}$. \\

\paragraph{Case III: $m_{4} \sim m_{1}$.} In this case, where the fourth mass eigenstate is quasi-degenerate with the first one, at first approximation we can set $\Delta m_{41}^{2} = \Delta m_{42}^{2}= 0$, and $\Delta m_{43}^{2} \approx - \Delta m_{31}^{2}$. For ($\theta_{14} \neq 0, \theta_{24} = 0$) case, we have  
\begin{eqnarray}
P(\nu_{e}\rightarrow \nu_{e}) = 1 - \left[\cos^{4}\theta_{14}\sin^{2}(2\theta_{{13}}) + \sin^{2}(2\theta_{14})\sin^{2}\theta_{13} \right]\sin^{2}\left(\frac{\Delta m^{2}_{31} L}{4E_\nu}\right)~,  
\label{eq:Pee_III_th24_0}
\end{eqnarray}
and for ($\theta_{14}=0, \theta_{24} \neq 0$), we obtain
\begin{eqnarray}
P(\nu_{\mu}\rightarrow \nu_{\mu}) = 1 - 4\left[\cos^{2}\theta_{24}\cos^{2}\theta_{13}\sin^{2}\theta_{{23}}(1 - \cos^{2}\theta_{24}\cos^{2}\theta_{13}\sin^{2}\theta_{{23}})  \right]\sin^{2}\left(\frac{\Delta m^{2}_{31} L}{4E_\nu}\right)~. 
\label{eq:Pmumu_III_th14_0}
\end{eqnarray}

Equations (\ref{eq:Pee_III_th24_0}) and (\ref{eq:Pmumu_III_th14_0}) clearly show that both $\theta_{14}$ and $\theta_{24}$ can be probed in survival channels, although weaker bounds can be derived on $\theta_{14}$ due to the $\sin^2\theta_{13}$ factor in Eq.~(\ref{eq:Pee_III_th24_0}). In addition, these limits are valid for any $\Delta m_{41}^2\lesssim10^{-4}~{\rm eV}^2$, as they originate from $\Delta m_{31}^2$-induced oscillations.

To illustrate some of the features discussed in Eqs.~(\ref{eq:Pmue_th24_0})-(\ref{eq:Pmumu_III_th14_0}), in Figure \ref{fig:probBeam} we show the $\nu_e$-appearance and $\nu_\mu$-survival oscillation probabilities, respectively, in the left and right panels, obtained by numerical calculation without using approximated formulas. The upper panels display the oscillation probabilities, and the lower panels illustrate the difference in probabilities between the $(3+1)$ scenario and the standard $3\nu$ scheme. The $\Delta m_{41}^2$-independent effect of $\theta_{14}$ on $\nu_e$-appearance can be seen by comparing the solid red and dashed violet curves in the left panel. The effect of $\theta_{24}$ on $P(\nu_\mu\to\nu_e)$ is exactly the same as the effect of $\theta_{14}$. The larger effect of $\theta_{24}$ on $\nu_\mu$-survival probability can be witnessed in the right panel (compare the mixing angle values used in the left and right panels), especially by looking at the red and blue curves which are manifestation of Case II and Case III. In contrast to $P(\nu_\mu\to\nu_e)$ in Eqs.~(\ref{eq:Pmue_th24_0}) and (\ref{eq:Pmue_th14_0}), there are terms in $P(\nu_\mu\to\nu_\mu)$ in Eqs.~(\ref{eq:Pmumu_I_th14_0}), (\ref{eq:Pmumu_II_th14_0}) and (\ref{eq:Pmumu_III_th14_0}) that are not suppressed by the $\sin^2\theta_{13}$ factor. 

\subsection{Atmospheric Neutrinos\label{sec:pheno_atm}}

\begin{figure}[ht]
    \centering
    \includegraphics[scale = 0.98]{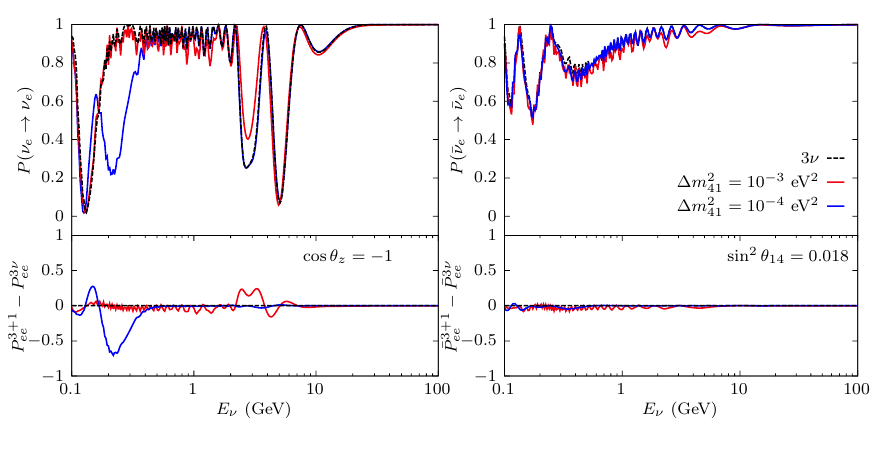}
    \includegraphics[scale = 0.98]{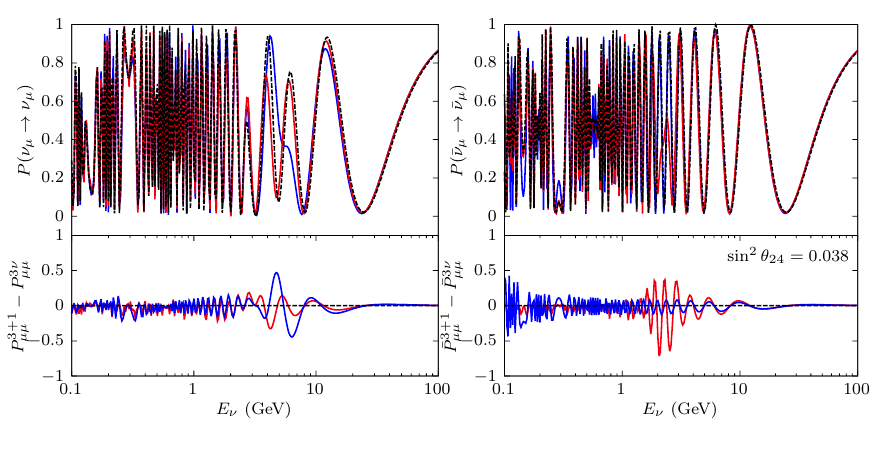}
    \caption{Electron (top panels) and muon (bottom panels) neutrino survival probabilities, $P_{ee}$ and $P_{\mu\mu}$, as functions of neutrino energy $E_{\nu}$ for up-going trajectories ($\cos \theta_{z} = -1$). Left panels correspond to neutrinos, and right panels to antineutrinos. The dashed black line shows the standard $3\nu$ scheme, while solid colored curves correspond to the (3+1) sterile neutrino scenario. In the top part of each panel we show survival probabilities for two representative sterile mass-squared splittings, $\Delta m^2_{41} = 10^{-4}\,\mathrm{eV}^2$ and $10^{-3}\,\mathrm{eV}^2$, compared with the standard case, while the bottom part shows the differences between the sterile and standard probabilities, $P_{\alpha\beta}^{3+1}-P_{\alpha\beta}^{3\nu}$. Normal ordering among active neutrinos was assumed. We note that as done for Figure~\ref{fig:probBeam}, all curves were obtained by numerical computations without using approximated probability formulas found in the previous subsection.}
    \label{fig:probAtm}
\end{figure}

Beyond accelerator neutrinos, atmospheric neutrinos provide a natural laboratory for probing sterile neutrino oscillations over long baselines and in the presence of matter effects. In contrast to the discussions presented in the previous subsection, for atmospheric neutrinos the effect of sterile neutrinos originates usually from resonant conversions induced by matter effect. Along the propagation through Earth, atmospheric neutrinos experience coherent interactions with matter that can resonantly enhance the active-sterile transitions via MSW~\cite{Mikheyev:1985zog,Nunokawa:2003ep,Choubey:2007ji,Esmaili:2012nz,Esmaili:2013cja,Esmaili:2013vza} and parametric~\cite{Liu:1997yb,Liu:1998nb} resonances. 

Figure~\ref{fig:probAtm} shows these effects for up-going trajectories ($\cos \theta_z =-1$), where the upper and lower panels show, respectively, the electron and muon neutrino survival probabilities, $P_{ee}$ and $P_{\mu\mu}$, for two representative values of $\Delta m_{41}^2$, and for neutrinos and antineutrinos. The dashed black curves denote the standard $3\nu$ scheme, while the solid colored curves correspond to the $(3+1)$ scenario. In each panel, the lower part shows the difference $P_{\alpha\beta}^{3+1}-P_{\alpha\beta}^{3\nu}$, highlighting the distortions introduced by sterile mixing. For the electron neutrino, the impact of the sterile state is most visible at sub-GeV energies (for the chosen $\Delta m_{41}^2$ values), where the resonance feature arises from the matter-induced potential difference between the electron and sterile neutrinos, effectively given by $V_{\mathrm{eff}}=V_{\rm CC}-V_{\rm NC} \simeq V_{\rm CC}/2$. 

For $\nu_e\to\nu_e$ channel resonance dip appears at $E_\nu \simeq 0.3$ GeV  for $\Delta m_{41}^2 = 10^{-4}\; \mathrm{eV}^2$, a in, consistent with the resonance condition $2E_\nu V_{\mathrm{eff}}\simeq \Delta m_{41}^2 \cos 2\theta_{14}$~\cite{Esmaili:2013cja}. At larger splittings, $\Delta m_{41}^2 = 10^{-3}\,\mathrm{eV}^2$, the resonance moves to higher energies, producing a smaller but still noticeable distortion in $\nu_e$ survival probability near $3$~GeV. For comparison, the standard atmospheric 1-3 resonance occurs when $V_{\mathrm{eff}}=V_{\rm CC}$. Since the electron-sterile system experiences a smaller effective potential ($V_{\mathrm{eff}}\simeq V_{\rm CC}/2$), and the resonance energy scales as $E_\nu \propto 1/V_{\mathrm{eff}}$, the corresponding electron-sterile resonance occurs at a higher energy than the 1-3 resonance. 

For the muon channel, shown in the lower panels of Figure~\ref{fig:probAtm}, both the standard and sterile neutrino scenarios display rapid oscillations at sub-GeV energies. The distortions at multi-GeV scales arise from matter effects in the $\nu_\mu \to \nu_s$ system, where the effective potential is given by $V_{\mathrm{eff}}=V_{\mu s}=-V_{\rm NC}$. For $\Delta m_{41}^2 = 10^{-4}\,\mathrm{eV}^2$, a resonance emerges at $E_\nu \simeq 6$~GeV, consistent with the condition $2E_\nu V_{\mathrm{eff}}\simeq \Delta m_{43}^2 \cos 2\theta_{24}$, with $\Delta m_{43}^2 \equiv \Delta m_{41}^2 - \Delta m_{31}^2$. Since $V_{\mathrm{eff}}<0$ for muon-sterile neutrino system, the resonance appears in the $\nu_\mu$ channel when $\Delta m_{41}^2 < \Delta m_{31}^2$ ($\Delta m_{43}^2<0$), while for $\Delta m_{41}^2 > \Delta m_{31}^2$ ({\it i.e.}, $\Delta m_{4i}^2>0$ for $i=1,2,3$) it occurs in $\overline{\nu}_\mu$. In the limit $m_4 \to m_1$ (or $\Delta m_{41}^2\to 0$), where $\Delta m_{43}^2\to -\Delta m_{31}^2$, the resonance remains at $E_\nu\sim6$~GeV in $\nu_\mu\to\nu_s$ channel. For larger sterile masses, $\Delta m_{41}^2 = 10^{-3}\,\mathrm{eV}^2$, the resonance at $\sim3$~GeV in $\bar{\nu}_\mu\to\bar{\nu}_s$ is the widely discussed resonance in the literature~\cite{Nunokawa:2003ep,Esmaili:2012nz,Esmaili:2013vza,Esmaili:2013cja}, which especially led to strong constraints on $(3+1)$ scenario from IceCube data~\cite{IceCube:2017ivd,IceCube:2016rnb,IceCubeCollaboration:2024nle} for $\Delta m_{41}^2\sim\mathcal{O}(1)~{\rm eV}^2$.

Within the Hyper-K energy window $(0.1\lesssim E_\nu /\mathrm{GeV}\lesssim 100)$, using atmospheric neutrinos, we probe the active-sterile resonances in both $\nu_e$($\bar{\nu}_e$) and $\nu_{\mu}$($\bar{\nu}_\mu$) channels. Also, we set $\theta_{34}=0$, since nonzero values of $\theta_{34}$ would generally strengthen the constraints on other active-sterile mixing angles~\cite{Esmaili:2013vza}. Based on the oscillation probability curves in Figure~\ref{fig:probAtm}, Hyper-K is expected to complement the existing bounds on active-sterile mixing angles, particularly for the sub-eV values of $\Delta m_{41}^2$.

\section{Accelerator Neutrino Analysis}\label{sec:beam}

Using the accelerator neutrinos, our goal is to explore the projections of the $(3+1)$ scenario's parameter space on the $(\sin^{2} \theta_{14}, \Delta m^{2}_{41})$ and $(\sin^{2} \theta_{24}, \Delta m^{2}_{41})$ planes, finding the regions that can be excluded by Hyper-K if the collected data are consistent with the standard $3\nu$ oscillation pattern. To accomplish this, we employ the $\chi^{2}$ function defined as
\begin{equation}
\begin{split}
\chi^2_{\mathrm{total}} (\vec{\Theta}) =\; \underset{\eta,\xi}{\mathrm{min}} & \left[  \sum_k^{\mathrm{bins}} \frac{\left(N_k^{\text{true}} - \bar{N}_{k}^{\mathrm{fit}}(\vec{\Theta}, \eta, \xi)\right)^2}{N_k^{\text{true}}}  + \sum_{s} \left(\frac{\eta^{s}}{\sigma_{\eta,s}}\right)^{2} + \sum_{b} \left(\frac{\eta^{b}}{\sigma_{\eta,b}}\right)^{2} \right.\\
& \left. + \sum_{k} \left(\frac{\xi_{k}^{s}}{\sigma_{\xi,s}}\right)^{2} + \sum_{k} \left(\frac{\xi_{k}^{b}}{\sigma_{\xi,b}}\right)^{2}\right]~,
\label{eq:chi2_acc}
\end{split}
\end{equation}
where 
\begin{equation}
\bar{N}_{k}^{\mathrm{fit}}(\vec{\Theta}, \eta, \xi) = \left(1 + \eta^{s} + \xi_{k}^{s}\right) S_k^{\text{fit}}(\vec{\Theta}) +\left(1 + \eta^{b} + \xi_{k}^{b}\right) B_k^{\text{fit}}(\vec{\Theta})~,
\end{equation}
with $N^{\mathrm{true}}_k$ and $\bar{N}^{\mathrm{fit}}_k$ representing the number of events in the $k$-th reconstructed energy bin associated with the standard $3\nu$ hypothesis and the $(3+1)$ test hypothesis, respectively. The index $k$ runs over all the reconstructed energy bins including both muon and electron neutrino samples, and both neutrino and antineutrino modes, corresponding to 34 bins (16 bins for $\nu_e$ and $\bar{\nu}_e$ channels, and 18 bins for $\nu_\mu$ and $\bar{\nu}_\mu$ channels). In $\bar{N}^{\mathrm{fit}}_k$, the signal $S_k^{\text{fit}}$ and background $B_k^{\text{fit}}$ components contribute to the event rate in the bin, which depend on the set of mixing parameters denoted by $\vec{\Theta}$. The systematic uncertainties are taken into account by the nuisance parameters $\eta$ and $\xi$, separately for signal and background components, labeled respectively by $s$ and $b$ superscripts. The parameter $\eta$ represents a flux normalization uncertainty, while $\xi_{k}$ represents the shape uncertainty in the $k$-th energy bin. The assumed prior uncertainties of flux normalization and shape are denoted respectively by $\sigma_{\eta}$ and $\sigma_{\xi}$, with the additional subscripts $s$ and $b$ for signal and background event rates.

The event rates in energy bins were obtained using the GLoBES library~\cite{Huber_2005,Huber_2007}, where the code has been modified to incorporate the sterile neutrino state. Our simulation of Hyper-K is based on the description provided in Ref.~\cite{Hyper-Kamiokande:2018ofw}. We assume ten years of collected data, with the ratio of exposures for neutrino and antineutrino modes as $\nu:\bar{\nu} = 1:3$, considering the appearance ($\nu_{\mu}\rightarrow \nu_{e}$) and disappearance ($\nu_{\mu}\rightarrow \nu_{\mu}$) channels. For appearance channel, the energy ranges from 0.045~GeV to 1.2~GeV, and for disappearance from 0.1~GeV to 2.7~GeV. The beam contamination ($\nu_{e}\rightarrow\nu_{e},\bar{\nu}_{e}\rightarrow\bar{\nu}_{e}$), which constitutes a background for the appearance channel, can significantly affect the exclusion bounds in the ($\sin^{2}\theta_{14}, \Delta m^{2}_{41}$) plane, and thus we only use it in the analysis searching for $\theta_{14}$. Throughout accelerator analysis, we have assumed 2.5\% normalization and 3\% shape uncertainty for $\nu_{e}$ ($\bar{\nu}_{e}$)-like events and 2\% normalization and 3\% shape uncertainty for $\nu_{\mu}$($\bar{\nu}_{\mu}$)-like events \cite{Xie:2023}. For beam contamination, we set 5\% for normalization and 3\% for shape uncertainty. The standard oscillation parameters $\theta_{23}$ and $\delta_{CP}$ are varied over the range defined in Table \ref{tab:std_osc},  with no penalty term, while all other standard parameters are kept fixed at their best-fit values, see Table~\ref{tab:std_osc}. After marginalization over the nuisance and standard oscillation parameters, the constraints on $\vec{\Theta} = (\sin^{2} \theta_{14}, \Delta m^{2}_{41})$ and $(\sin^{2} \theta_{24}, \Delta m^{2}_{41})$ at 95\% CL can be derived by the standard relation $\Delta\chi^2(\vec{\Theta}) = \chi^2(\vec{\Theta}) - \chi^2_{\rm min}\leq5.99$ (since we are performing a sensitivity analysis, in fact $\chi^2_{\rm min}=0$). Throughout the analysis, any mixing angle or phase related to the 4th mass eigenstate not explicitly included in $\vec{\Theta}$ was set to zero. Normal ordering was assumed throughout the article, unless otherwise noted. Finally, we do not explicitly simulate the near detector of Hyper-K, however, the values of systematic uncertainties used during the analysis already account for the impact of the near detector.

\renewcommand\arraystretch{1.5}
\renewcommand\tabcolsep{0.15cm}
\begin{table}[ht]
\caption{Standard oscillation parameters used throughout the accelerator analysis~\cite{Esteban:2024eli,PDGNavas2024,deSalasGlobalFit2020}, and their range for marginalization.}
\centering
\begin{tabular}{cccc}
\hline
\hline
\textbf{Parameter} & \textbf{Value} & \textbf{Treatment} & \textbf{Range} \\ 
\hline
\hline
$\sin^{2} \theta_{12}$ & 0.308 & fixed & -- \\
$\sin^{2} \theta_{13}$ & 0.02215 & fixed & --\\
$\sin^{2} \theta_{23}$ & 0.472 & marginalized & [0.435, 0.585] \\
$\delta_{CP}$ & $\pi/2$ & marginalized & $[-\pi,\pi]$\\
$\Delta m_{21}^{2}$ & $7.49\times10^{-5}$ eV$^2$ & fixed & -- \\
$\Delta m_{31}^{2}$ & $2.513\times10^{-3}$ eV$^2$ & fixed & -- \\
\hline
\hline
\end{tabular}
\label{tab:std_osc}
\end{table}

\begin{figure}[ht]
    \centering
    \includegraphics[width=1.0\linewidth]{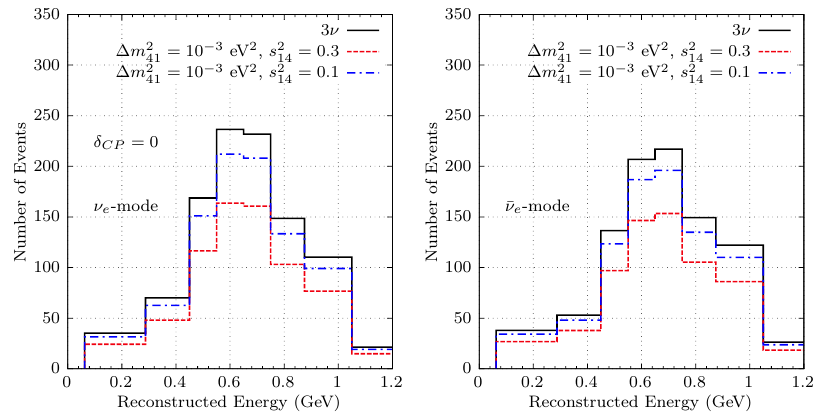}
    \caption{Number of expected events in the appearance channel assuming ten years of collected data by Hyper-K experiment. The left (right) panel displays the event's energy distribution for neutrino (antineutrino) mode. We compare the $3\nu$, depicted by black solid line, with the $(3+1)$ scenario assuming $\Delta m_{41}^2 = 10^{-3}~{\rm eV}^2$ and $\sin^2\theta_{14}=0.3$ (red dashed line) and $\sin^2\theta_{14}=0.1$ (blue dot-dashed line). For standard oscillation parameters not labeled in the figure, we have assumed the values reported in Table~\ref{tab:std_osc}. 
    The mixing parameters involving the 4th state not displayed in the figure are set to zero.}
    \label{fig:events_nue}
\end{figure}

\begin{figure}[H]
    \centering
    \includegraphics[width=1.0\linewidth]{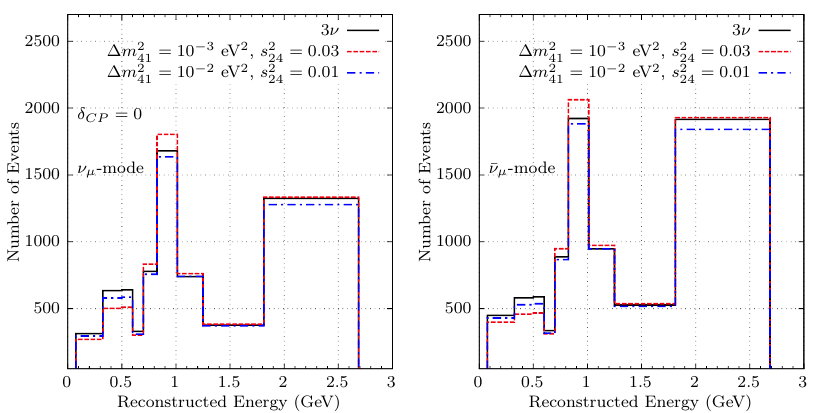}
    \caption{The same as Figure~\ref{fig:events_nue}, but for disappearance channel, which can probe the $(\sin^2\theta_{24},\Delta m_{41}^2)$ parameters in $(3+1)$ scenario. $(\sin^2\theta_{24},\Delta m_{41}^2)=(0.03,10^{-3}~{\rm eV}^2)$ is depicted by red dashed line, and for $(\sin^2\theta_{24},\Delta m_{41}^2)=(0.01,10^{-2}~{\rm eV}^2)$ we use the blue dot-dashed line.}
    \label{fig:events_numu}
\end{figure}

For illustrative purposes, Figures~\ref{fig:events_nue} and \ref{fig:events_numu} show the energy distribution of events at Hyper-K, for ten years of data collection, for $(\sin^{2} \theta_{14}, \Delta m^{2}_{41})$ and $(\sin^{2} \theta_{24}, \Delta m^{2}_{41})$ analyses, respectively. We compare the $3\nu$ and $(3 + 1)$ scenarios for a set of sterile neutrino parameters relevant to each channel and to which Hyper-K will be sensitive. Figure~\ref{fig:events_nue} illustrates the appearance channels for neutrino (left) and antineutrino (right) modes, while  Figure~\ref{fig:events_numu} presents the disappearance channels for neutrino (left) and antineutrino (right) modes. The black solid line represents the standard three neutrino case, while colored lines illustrate the (3+1) scenario. 

Considering the observed event rate in the muon neutrino survival channel (Figure \ref{fig:events_numu}), the set of parameters $(\sin^2\theta_{24},\Delta m_{41}^2)=(0.03, 10^{-3} \text{ eV}^2)$ results in a significant modification of the event rate compared to the standard three-flavor case. This modification has a distinct spectral shape that cannot be mimicked by a simple adjustment of the flux normalization. Consequently, the experiment can strongly exclude this specific parameter combination.  In contrast, for the parameter set $(\sin^2\theta_{24},\Delta m_{41}^2)=(0.01, 10^{-2} \text{ eV}^2)$ in the muon survival channel, or for $(\sin^2\theta_{14},\Delta m_{41}^2)$ in the electron neutrino appearance channel,  the predicted event rates are more degenerate with the impact of systematic uncertainties. Therefore, constraining these values of the parameters is primarily limited by systematic errors.


\section{Atmospheric Neutrino Analysis\label{sec:atm}}

Hyper-K is expected to accumulate a substantial amount of atmospheric neutrino data during its operational period, covering energies from approximately $100$~MeV to $100$~GeV. Although Hyper-K will benefit from improved momentum and directional reconstruction capabilities for leptons (electrons and muons) generated in neutrino interactions, our analysis conservatively models Hyper-K as a scaled-up Super-K detector, adopting identical detection efficiencies and performance characteristics. More precisely, we assume a fiducial volume of 187 kton for the Hyper-K tank~\cite{Hyper-Kamiokande:2018ofw}, which is approximately 8.3 times the fiducial volume of Super-K.

In our analysis, following the approach used in Super-K and explained comprehensively in~\cite{Gonzalez-Garcia:2007dlo}, we categorize neutrino events into two classes: -- Fully Contained (FC) events, occur when the charged lepton (electron or muon) produced by the neutrino interaction remains fully within the detector volume; -- Partially contained (PC) events, occur when the muon produced by a $\nu_\mu/\bar{\nu}_\mu$ interaction originates within the detector but subsequently exits its volume. Electron-like PC events do not occur. We further subdivide FC events into three groups based on the reconstructed lepton momentum $p_{l}$: MultiGeV ($p_{l} > 1.2$~GeV), SubGeVlow ($100~{\rm MeV}<p_{l}<400$~MeV) and SubGeVhigh ($400~{\rm MeV}<p_{l}<1.2$~GeV) events. No momentum cut is applied to PC events.   

The number of events in the $i$-th bin of zenith angle, for a set of mixing parameters $\vec{\Theta}$, is computed as~\cite{Gonzalez-Garcia:2007dlo} 
\begin{equation}\label{eq:atmevent}
    \begin{split}
    N_i (\vec{\Theta}) &= n_t T \sum_{\alpha,\beta,\pm} \int_{0}^{\infty} {\rm d}h \int_{-1}^{1} {\rm d}\cos\theta_\nu \int_{E_{\rm min}}^{\infty} {\rm d} E_\nu
    \int_{E_{\rm min}}^{E_\nu} {\rm d}E_l \int_{-1}^{1} {\rm d}\cos\theta_a \int_{0}^{2\pi} {\rm d}\phi_a \\
    &\quad \frac{{\rm d}^3\Phi^{\pm}_{\alpha}}{{\rm d}E_\nu {\rm d}\cos\theta_\nu {\rm d}h}\left(E_\nu,\cos\theta_\nu,h\right) \times P^{\pm}_{\alpha\to\beta} (E_\nu,\cos\theta_\nu,h|\vec{\Theta}) \times \\
    &\quad \left[ \frac{{\rm d}^2\sigma_{\beta}^{\pm}}{{\rm d}E_l{\rm d}\cos\theta_a} \pi_{\rm ring}\right] (E_\nu,E_l,\cos\theta_a) \times\epsilon_{\beta}^{\rm bin} \left(E_l, \cos\theta_l\left(\cos\theta_\nu,\cos\theta_a,\phi_a\right)\right)~,
    \end{split}
\end{equation}
where $n_t$ is the number of target nuclei, and $T$ denotes the Hyper-K operational time. The indices $+(-)$ correspond to neutrinos (antineutrinos), while $\alpha$ and $\beta$ indicate neutrino flavors ($\nu_e$ and $\nu_\mu$). The atmospheric neutrino triple-differential flux, expressed as a function of the neutrino energy $E_\nu$, the zenith angle $\theta_\nu$, and the neutrino production altitude $h$, is denoted by ${\rm d}^3\Phi^{\pm}{\alpha}/{\rm d}E_\nu {\rm d}\cos\theta_\nu {\rm d}h$, which we use the HKKM2014 model~\cite{Honda:2015fha}. To reduce computational load, we fix the neutrino production altitude to $h=20$~km. This assumption does not affect $\Delta m_{41}^2\lesssim10^{-2}~{\rm eV}^2$, for $E_\nu\simeq1$~GeV. For $\Delta m_{41}^2\gtrsim10^{-2}~{\rm eV}^2$, the altitude distribution can affect the active-sterile oscillation for down-going events ($\cos\theta_\nu\simeq1$) and thus, its omission lead to conservative results for these values of $\Delta m_{41}^2$. The atmospheric neutrino flux at low energies depends on solar activity and is defined as $\Phi = c_{\rm min} \Phi_{\rm min} + c_{\rm max} \Phi_{\rm max}$, with fluxes $\Phi_{\rm min}$ and $\Phi_{\rm max}$ representing minimum and maximum solar activities, respectively. The weighing factors $c_{\rm min}$ and $c_{\rm max}=1-c_{\rm min}$ depend on the data collection period, and we use $c_{\rm min}=0.47$ as a nominal value. 

The oscillation probabilities for $\nu_\alpha \to \nu_\beta$ ($\bar{\nu}_\alpha \to \bar{\nu}_\beta$) channels are denoted by $P^{+}_{\alpha\to\beta}$ ($P^{-}_{\alpha\to\beta}$). The differential neutrino-water cross section, with respect to the produced lepton (either electron or muon) energy $E_l$, and the opening angle $\theta_a$ (the angle between the produced lepton and the incoming neutrino), can be written as
 \begin{equation}
     \frac{{\rm d}^2\sigma_{\beta}^{\pm}}{{\rm d}E_l{\rm d}\cos\theta_a} \eta_{\rm ring} = \frac{{\rm d}^2\sigma_{{\rm CCQE},\beta}^{\pm}}{{\rm d}E_l{\rm d}\cos\theta_a} \eta^{\rm CCQE}_{\rm ring} + \frac{{\rm d}^2\sigma_{{\rm CC1}\pi,\beta}^{\pm}}{{\rm d}E_l{\rm d}\cos\theta_a} \eta^{{\rm CC1}\pi}_{\rm ring} + \frac{{\rm d}^2\sigma_{{\rm CC}N\pi,\beta}^{\pm}}{{\rm d}E_l{\rm d}\cos\theta_a} \eta^{{\rm CC}N\pi}_{\rm ring} + \frac{{\rm d}^2\sigma_{{\rm DIS},\beta}^{\pm}}{{\rm d}E_l{\rm d}\cos\theta_a} \eta^{\rm DIS}_{\rm ring}~,
    \end{equation}   
where the terms in the correspond to Charged Current Quasi-Elastic (CCQE), Charged Current one $\pi$ production (CC$1\pi$), Charged Current $N\pi$ production (CC$N\pi$, $N\ge2$), and Deep Inelastic Scattering (DIS) interactions. The $\eta_{\rm ring}$ factors take into account the acceptance of the detector for single-ring tagging, which are different from one just for FC events (as in Super-K, no ring-tagging is used for PC events). As in~\cite{Gonzalez-Garcia:2007dlo}, we use $\eta^{\rm CCQE}_{\rm ring}=0.96/(1 + 0.03p_l^{1.38})$ for electron-like events (where $p_l$ is the lepton momentum in GeV) and $\eta^{\rm CCQE}_{\rm ring}=0.97$ for muon-like events. $\eta^{{\rm CC1}\pi}_{\rm ring}$ is the probability that produced pion's energy is less than $2.1 m_\pi$, and thus does not produce an extra ring ($m_\pi$ is the pion mass). The $\eta^{{\rm CC1}\pi}_{\rm ring}$ is calculated from the differential cross section ${\rm d}^2\sigma_{{\rm CC}1\pi,\beta}^{\pm}/{\rm d}E_\pi{\rm d}\cos\theta_a$, where $E_\pi$ is the pion energy. $\eta^{{\rm CC}N\pi}_{\rm ring}$ is the probability that the total energy of produced pions is less than $2.1 n_\pi m_\pi$, where $n_\pi$ is the mean multiplicity of pion. The $\eta^{{\rm CC}N\pi}_{\rm ring}$ and $n_\pi$ are computed from the differential cross section ${\rm d}^2\sigma_{{\rm CC}N\pi,\beta}^{\pm}/{\rm d}E_{N\pi}{\rm d}\cos\theta_a$, where $E_{N\pi}$ is the total energy of produced pions. Finally, since we have separated the CC$N\pi$ from DIS, we set $\eta^{{\rm DIS}}_{\rm ring}=1$. All differential cross sections of neutrino-water interactions are generated by the NEUT~\cite{Hayato:2021heg}. 

The detector efficiency $\epsilon_\beta^{\rm bin}$, a function of lepton energy $E_l$ and zenith angle of the lepton $\theta_l$ (given by $\cos\theta_l = \cos\theta_\nu \cos\theta_a - \sin\theta_\nu \sin\theta_a \cos\phi_a$, where $\phi_a$ is the azimuthal opening angle), decomposes as: 
\begin{equation}
\epsilon_\beta^{\rm bin} (E_l,\cos\theta_l) = \epsilon_\beta^{\rm thr} (E_l) \times \epsilon_\beta^{\rm con} (E_l,\cos\theta_l) \times \epsilon_{\rm bin}^{\rm zen} (\cos\theta_l)~.
\end{equation}
Here, $\epsilon_\beta^{\rm thr}(E_l)$ applies the momentum cut for FC events, while no cuts are applied for PC events. We assume the same resolution in momentum measurement as the Super-K, that is $0.6\% + 2.6\%/\sqrt{p_l/{\rm GeV}}$ for single-ring electrons and $1.7\% + 0.7\%/\sqrt{p_l/{\rm GeV}}$ for single-ring muons. The $\epsilon_\beta^{\rm con} (E_l,\cos\theta_l)$ is the probability that a muon with energy $E_l$ and direction $\cos\theta_l$ contributes to the PC or FC events. We computed it using the muon range in water~\cite{Lipari:1991ut} and via a Monte Carlo simulation. Lastly, $\epsilon_{\rm bin}^{\rm zen} (\cos\theta_l)$ represents the probability that a lepton from direction $\cos\theta_l$ contributes to the bin under consideration. The following resolutions in measuring the direction have been taken into account: $3.0^\circ$ for single-ring electron-like events, $1.8^\circ$ for single-ring FC muon-like events, and $2.8^\circ$ for PC muon-like events.  

Different data sets of atmospheric neutrinos with distinct event topologies provide a handle on probing different regions of the sterile neutrino parameter space. We use the MultiGeV($e$), SubGeVhigh($e$) and SubGeVlow($e$) events to probe $\vec{\Theta}=(\theta_{14},\Delta m_{41}^2)$, and MultiGeV(PC-$\mu$), MultiGeV(FC-$\mu$) and SubGeVhigh($\mu$) to probe $\vec{\Theta}=(\theta_{24},\Delta m_{41}^2)$ in $(3+1)$ scenario. To estimate the sensitivity of Hyper-K to sterile neutrino scenario, for each event's type we use the following $\chi^2$ function 
\begin{equation}
\chi^2_{\mathrm{total}} (\vec{\Theta}) = \underset{\{f_i\}}{\mathrm{min}}\left[ \sum_k^{\mathrm{bins}} \frac{\left(N_k^{\text{true}} - \bar{N}_{k}^{\mathrm{fit}}(\vec{\Theta})\right)^2}{N_k^{\text{true}}}  + \sum_{j=1}^{16} \left(\frac{f_j}{\sigma_j}\right)^{2}\right]~,
\label{eq:chi2_atm}
\end{equation}
where 
\begin{equation}
\bar{N}_{k}^{\mathrm{fit}} = \left(1 + \sum_{i=1}^{16} \pi_{ik} f_i\right) N_k^{\text{fit}}~, 
\end{equation}
with $N_k^{\text{true}}$ and $N_k^{\text{fit}}$ being the number of events in the $k$-th bin of zenith angle. $f_i$ denotes the nuisance parameters, totally 16, with the following uncertainties $\sigma_i$: atmospheric neutrino flux uncertainties (20\% normalization, 5\% energy spectrum tilt, 5\% up-down asymmetry, 5\% $\nu/\bar{\nu}$ ratio asymmetry, 2.5\% $\nu_\mu/\nu_e$ ratio), cross section uncertainties (15\% normalization of CCQE, CC1$\pi$ and CC$N\pi$+DIS, 0.5\% ratio of $\sigma_{\nu_\mu}/\sigma_{\nu_e}$ ratio for CCQE, CC1$\pi$ and CC$N\pi$+DIS interactions), and systematic uncertainties related to hadronic interactions simulation, particle identification, ring counting, fiducial volume and energy calibration. All the uncertainties and the corresponding values of $\pi_{ik}$ are taken from~\cite{Gonzalez-Garcia:2007dlo}.  

\begin{figure}[ht]
    \centering
    \includegraphics[width=0.49\linewidth]{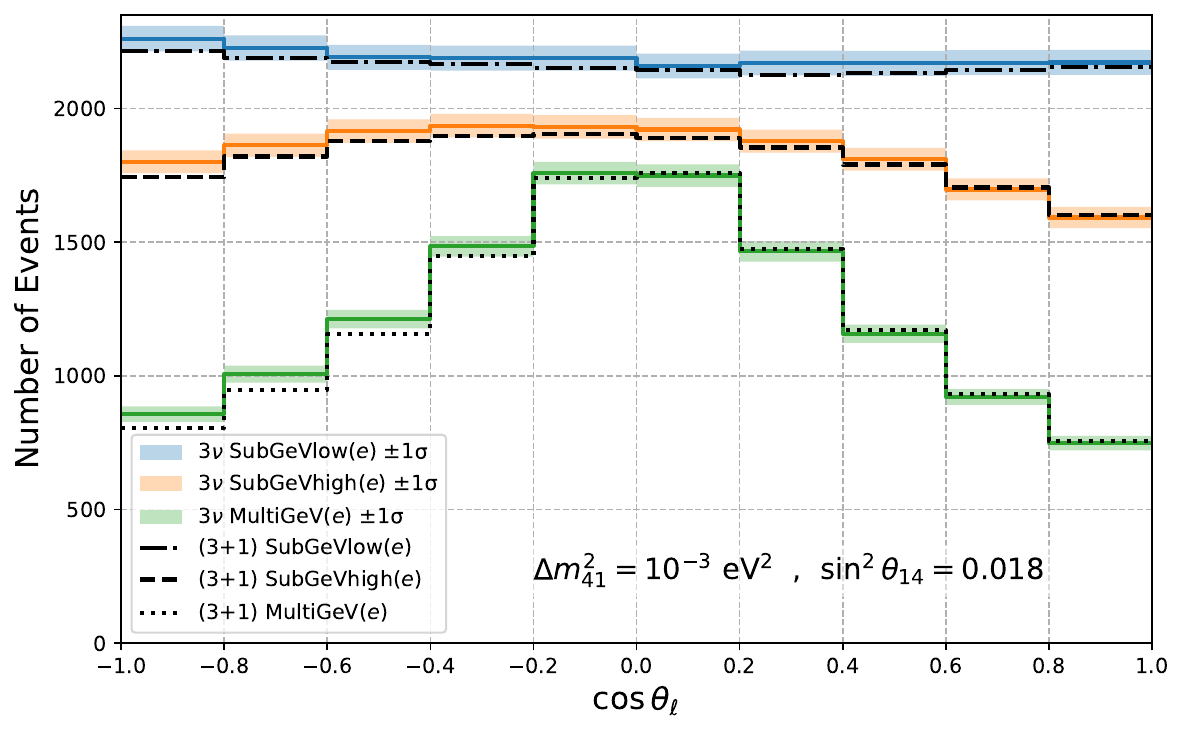}
    \includegraphics[width=0.49\linewidth]{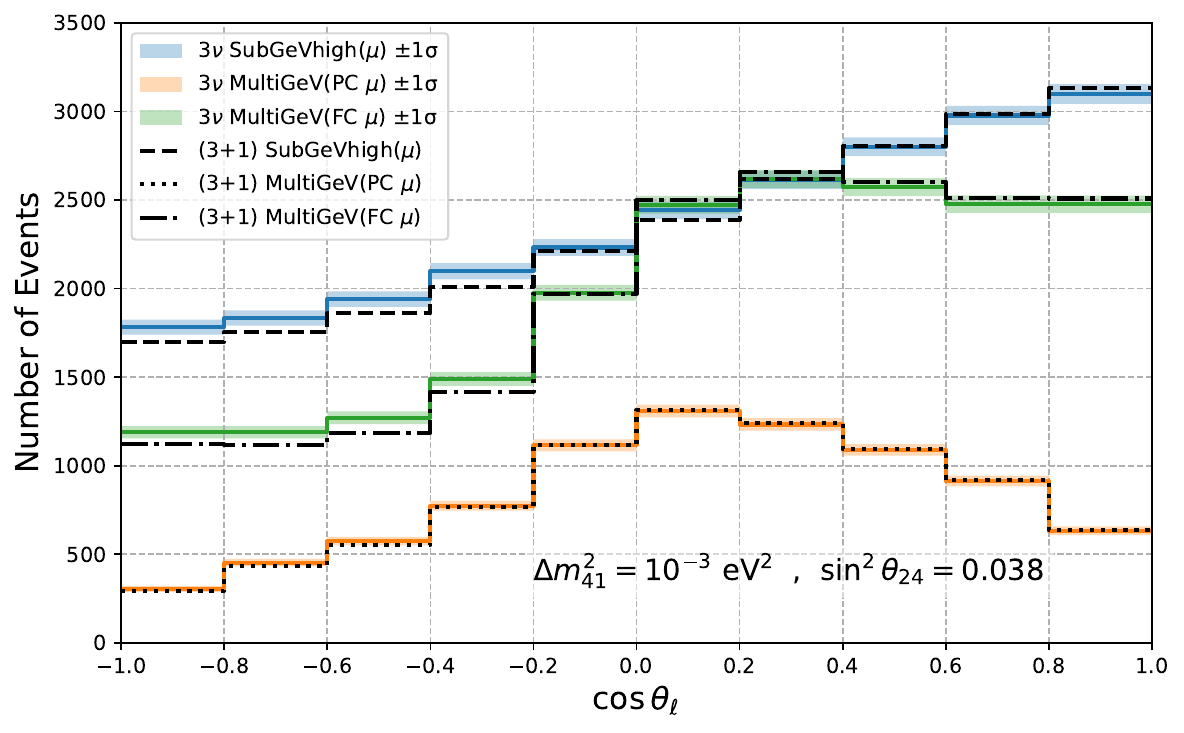}
    \caption{Atmospheric neutrino event distributions as function of the zenith angle of lepton $\theta_l$. {\textit{Left panel}}: Distributions of SubGeVlow($e$), SubGeVhigh($e$) and MultiGeV($e$) events, in the $3\nu$ and $(3+1)$ scenarios, whit active-sterile mixing values $(\sin^2\theta_{14},\Delta m_{41}^2)=(0.018,10^{-3}~{\rm eV}^2)$. {\textit{Right panel}}: Distributions of SubGeVhigh($\mu$), MultiGeV(PC $\mu$) and MultiGeV(FC $\mu$) events, in the $3\nu$ and $(3+1)$ scenarios, for $(\sin^2\theta_{24},\Delta m_{41}^2)=(0.038,10^{-3}~{\rm eV}^2)$. In both panels, the color-shaded bands show the $\pm1\sigma$ statistical uncertainty of $3\nu$ scenario. In the $(3+1)$ scenario, the event distributions are shown after marginalization over nuisance parameters. For both panels we assume 6 years of data collection. All the mixing parameters of active neutrinos are fixed to the values reported in Table~\ref{tab:std_osc}.}
    \label{fig:events_atm}
\end{figure}

Using the formalism outlined in this section, we have reproduced the expected zenith-angle distribution of atmospheric neutrinos for various event types in Hyper-K~\cite{Hyper-Kamiokande:2018ofw}. Figure~\ref{fig:events_atm} illustrates the impact of sterile neutrinos on these distributions. In the left panel, we compare the $3\nu$ scenario (solid curves) with the $(3+1)$ scenario (dash-dot, dashed and dotted curves, respectively corresponding to SubGeVlow($e$), SubGeVhigh($e$) and MultiGeV($e$) events) for parameter values $(\sin^2\theta_{14},\Delta m_{41}^2) = (0.018,10^{-3}~{\rm eV}^2)$. The shaded regions around the $3\nu$ curves indicate the $\pm1\sigma$ statistical uncertainties. In the right panel, we compare the zenith distributions of SubGeVhigh($\mu$), MultiGeV(PC $\mu$), and MultiGeV(FC $\mu$) events between the $3\nu$ and $(3+1)$ scenarios, where in the latter we assume $(\sin^2\theta_{24},\Delta m_{41}^2) = (0.038,10^{-3}~{\rm eV}^2)$. In both panels, the $(3+1)$ event distributions are obtained after marginalizing the $\chi^2$ over nuisance parameters. 

It is worth noting that in our atmospheric neutrino analysis, we adhere to the conventional event-type classifications used in Super-K (MultiGeV, SubGeVhigh, etc.). Although finer binning in lepton momentum is possible, given that momentum resolution is on the order of a few percent, we maintain these standard categories to allow direct comparison and to ensure conservative results.

\section{Results and Discussions}\label{sec:results}

\begin{figure}[h]
    \centering
    \includegraphics[scale = 1.]{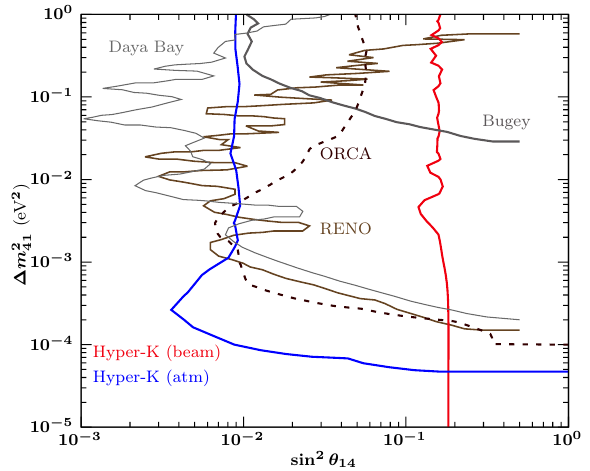}
    \caption{The expected 95\% CL exclusion curve in the ($\sin^{2}\theta_{14}, \Delta m^{2}_{41}$) plane assuming ten years of only beam-based events (red curve) and six years of only atmospheric-based events (blue curve) at Hyper-K. For comparison, we also show the limits from Daya Bay at 95\% CL (light-gray curve)~\cite{DayaBay_2016}, Bugey at 90\% CL (dark-gray curve)~\cite{Bugey_1994}, and RENO at 95\% CL (brown line)~\cite{RENO_2020}. The projected sensitivity of KM3NET/ORCA at 95\% CL~\cite{KM3NeT:2021uez} is depicted by the dark-brown dashed curve. Normal mass ordering among active neutrinos is assumed.}
    \label{fig:th14dm41-beam}
\end{figure}

\begin{figure}[h]
    \centering
    \includegraphics[scale = 1.]{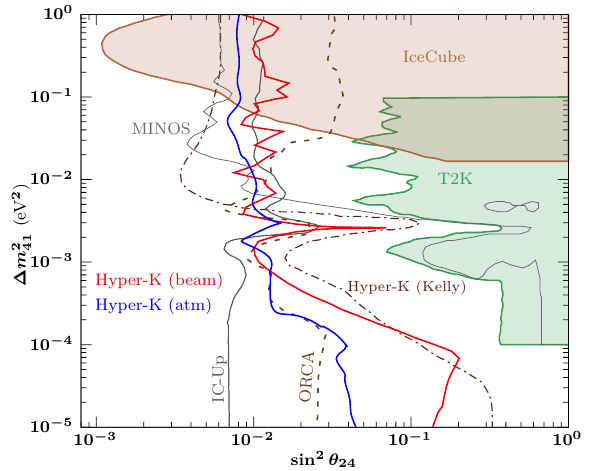}
    \caption{The expected 95\% CL exclusion curve in the ($\sin^{2}\theta_{24}, \Delta m^{2}_{41}$) plane assuming ten years of only beam-based events (red curve) and six years of only atmospheric-based events (blue curve) at Hyper-K. For comparison, we show the existing limits from MINOS/MINOS+ at 90\% CL~\cite{MINOS:2017cae} (light-gray curve), T2K at $2\sigma$ CL~\cite{deGouvea:2022kma}(green filled region), and IceCube at 99\% CL~~\cite{IceCubeCollaboration:2024nle} (light-brown filled region). The projected sensitivity of Hyper-K from~\cite{Kelly_2017} at 95\% CL, from the combined atmospheric and accelerator neutrino events, is shown by brown dot-dashed curve. The sensitivities of IceCube-Upgrade~\cite{cabrera2025} and KM3NET/ORCA~\cite{KM3NeT:2021uez}, at 90\% CL, are shown by dark-gray and light-brown-dashed curves, respectively. Normal mass ordering among active neutrinos is assumed.}
    \label{fig:th24dm41-beam}
\end{figure}

In this section, we present the results of minimizing the $\chi^{2}$ functions in Eqs.~(\ref{eq:chi2_acc}) and (\ref{eq:chi2_atm}), which consider the accelerator and atmospheric events at Hyper-K, respectively. The projection of Hyper-K sensitivity, at 95\% confidence level (CL), on the ($\sin^{2} \theta_{14}, \Delta m^{2}_{41}$) and ($\sin^{2} \theta_{24}, \Delta m^{2}_{41}$) planes are shown in Figs.~\ref{fig:th14dm41-beam} and \ref{fig:th24dm41-beam}, respectively.

The accelerator neutrinos at Hyper-K can constrain $\theta_{14}$ mainly through the modified $\nu_e\to\nu_e$ oscillation that is not suppressed by $\sin^2\theta_{13}$, which acts as background to the appearance signal channel $\nu_\mu\to\nu_e$ which its $\theta_{14}$-dependence is suppressed (see section~\ref{sec:pheno_acc}). The $\Delta m_{41}^2$-independent limit on $\theta_{14}$ from accelerator neutrinos, for Cases I and III of section~\ref{sec:pheno_acc}, that is $\Delta m_{41}^2\gtrsim10^{-2}~{\rm eV}^2$ and $\Delta m_{41}^2\lesssim10^{-3}~{\rm eV}^2$, can be verified by the red curve in Fig.~\Ref{fig:th14dm41-beam}. The only feature in the limit from accelerator neutrinos, as expected, is at $10^{-3} \lesssim \Delta m^{2}_{41}/{\rm eV}^2 \lesssim 10^{-2}$, which corresponds to Case II (see Eq.~(\ref{eq:Pee_II_th24_0})). For atmospheric neutrinos, the sensitivity originates from the resonance in $\nu_e\to\nu_s$ conversion. Looking at Fig.~\ref{fig:probAtm}, a stronger bound on $\theta_{14}$ is expected at $\Delta m_{41}^2\sim10^{-4}~{\rm eV^2}$ than $10^{-3}~{\rm eV}^2$, which can be verified by the blue curve in Fig.~\ref{fig:th14dm41-beam}. For $\Delta m_{41}^2\lesssim10^{-4}~{\rm eV}^2$ the resonance in $\nu_e\to\nu_s$ moves to energies $\lesssim100$~MeV and thus the sensitivity drops. For $\Delta m^{2}_{41} \gtrsim 10^{-2}~{\rm eV}^{2}$, where the fast oscillations due to $\theta_{14}\neq0$ appear in the energy range of Hyper-K, the limit from atmospheric neutrinos becomes $\Delta m_{41}^2$-independent. In Fig.~\ref{fig:th14dm41-beam}, we also show the existing bounds from Daya Bay~\cite{DayaBay_2016}, Bugey~\cite{Bugey_1994}, and RENO~\cite{RENO_2020} experiments, and the projected sensitivity of KM3NET/ORCA~\cite{KM3NeT:2021uez}. Clearly, Hyper-K can play an important role in establishing bounds on $\theta_{14}$, particularly at $\Delta m_{41}^2\lesssim10^{-3}~{\rm eV}^2$.

The limit on $\theta_{24}$ from accelerator neutrinos, shown by the red curve in Fig.~\ref{fig:th24dm41-beam}, is almost independent of $\Delta m_{41}^2$ value for $\Delta m_{41}^2\gtrsim10^{-2}~{\rm eV}^2$, as expected from Eq.~(\ref{eq:Pmumu_I_th14_0}) of Case I in section~\ref{sec:pheno_acc}. At $\Delta m_{41}^2\approx\Delta m_{31}^2$ the bound deteriorates, since the active-sterile oscillation pattern mimics the active-active pattern. At $\Delta m_{41}^2\lesssim10^{-4}~{\rm eV}^2$, the sensitivity originates from Eq.~(\ref{eq:Pmumu_III_th14_0}) which can constrain $\theta_{24}$ due to $\Delta m_{31}^2$-induced oscillations. The projected sensitivity from atmospheric neutrinos, shown by the blue curve, surpasses the one from accelerator neutrinos at $\Delta m_{41}^2\lesssim10^{-3}~{\rm eV}^2$ due to the resonance in $\nu_\mu$ channel originating from $\Delta m_{43}^2$ (see section~\ref{sec:pheno_atm}). At large mass splittings, $\Delta m_{41}^2\gtrsim10^{-1}~{\rm eV}^2$, where the resonance in $\bar{\nu}_\mu$ channel lands in $\sim$~TeV energies, the existing limit from IceCube~\cite{IceCubeCollaboration:2024nle} is stronger. For comparison, we also show the existing limits from MINOS/MINOS+~\cite{MINOS:2017cae} and T2K~\cite{deGouvea:2022kma}, and the projected sensitivities of KM3NET/ORCA~\cite{KM3NeT:2021uez} and IceCube-Upgrade~\cite{cabrera2025}. The dot-dashed curve in Fig.~\ref{fig:th24dm41-beam} shows the estimated sensitivity of Hyper-K, from the combined atmospheric and accelerator neutrino events, taken from~\cite{Kelly_2017}. Although some features in this curve agree with our result, our analysis predicts an improved sensitivity in ($\sin^{2} \theta_{24}, \Delta m^{2}_{41}$) plane, especially at $\Delta m_{41}^2 \lesssim10^{-3}~{\rm eV}^2$.

\section*{acknowledgment}
A.~E. thanks Jake McKean for his invaluable assistance with neutrino cross section computations. This work was in part supported by Brazilian Funding Agencies, CNPq and CAPES. E.~C. thanks the support from CNPq scholarship No. 140121/2022-6 and CAPES-PDSE scholarship No. 88881.982354/2024-01. A.~M.~T. thanks the support from CAPES scholarship No. 88887.479307/2020-00 and grant No. 312263/2021-0 and CNPq scholarship No. 166794/2022-8. 

\bibliography{sterile-light.bib}

\end{document}

%% file: Packages.tex
\usepackage{CJKutf8}

\usepackage{tikz}
\usepackage{adjustbox}
\usepackage{amsmath}
\usepackage{graphicx}
\usepackage{fullpage}
\usepackage{color}
\usepackage[colorlinks=true,allcolors=blue]{hyperref}
\usepackage[capitalise]{cleveref}
\usepackage[utf8]{inputenc}
\usepackage{siunitx}
\usepackage{array}
\input{units}
\usepackage{pifont}
\usepackage{enumitem}
\usepackage{url}
\usepackage{rotating}
\usepackage{epsfig,graphics,rotate}
\usepackage{flushend}
\usepackage{bm}
\usepackage{amssymb}
\usepackage{amsmath}
\usepackage{amsfonts}
\usepackage{gensymb}
\usepackage{booktabs}

\usepackage{microtype}
\usepackage{multirow, makecell}
\usepackage{lipsum}

\usepackage{xspace}
\usepackage{comment}
\usepackage{floatrow}
\usepackage{ragged2e}
\usepackage{siunitx}

\usepackage{nicefrac, xfrac}
\usepackage{mathtools} 
\usepackage{relsize}   

\usepackage{float}

%% file: units.tex
\DeclareSIUnit \s {\second}
\DeclareSIUnit \ns {\nano\second}
\DeclareSIUnit \mus {\micro\second}
\DeclareSIUnit \ms {\milli\second}
\DeclareSIUnit \MB {\mega\byte}
\DeclareSIUnit \GB {\giga\byte}
\DeclareSIUnit \TB {\tera\byte}
\DeclareSIUnit \PB {\peta\byte}
\DeclareSIUnit \Mbps {\mega\bit/\s}
\DeclareSIUnit \Gbps {\giga\bit/\s}
\DeclareSIUnit \Tbps {\tera\bit/\s}
\DeclareSIUnit \Pbps {\peta\bit/\s}
\DeclareSIUnit \kton {\kilo\tonne} 
\DeclareSIUnit \kt {\kilo\tonne}
\DeclareSIUnit \kty {\kilo\tonne-\year}
\DeclareSIUnit \Mt {\mega\tonne}
\DeclareSIUnit \eV {\electronvolt}
\DeclareSIUnit \keV {\kilo\electronvolt}
\DeclareSIUnit \MeV {\mega\electronvolt}
\DeclareSIUnit \GeV {\giga\electronvolt}
\DeclareSIUnit \TeV {\tera\electronvolt}
\DeclareSIUnit \PeV {\peta\electronvolt}
\DeclareSIUnit \EeV {\exa\electronvolt}
\DeclareSIUnit \sr {sr}
\DeclareSIUnit \m {\meter}
\DeclareSIUnit \cm {\centi\meter}
\DeclareSIUnit \nm {\nano\meter}
\DeclareSIUnit \in {\inchcommand}
\DeclareSIUnit \km {\kilo\meter}
\DeclareSIUnit \kV {\kilo\volt}
\DeclareSIUnit \kW {\kilo\watt}
\DeclareSIUnit \MW {\mega\watt}
\DeclareSIUnit \MHz {\mega\hertz}
\DeclareSIUnit \mrad {\milli\radian}
\DeclareSIUnit \year {years}
\DeclareSIUnit \POT {POT}
\DeclareSIUnit \sig {$\sigma$}
\DeclareSIUnit\parsec{pc}
\DeclareSIUnit\lightyear{ly}
\DeclareSIUnit\foot{ft}
\DeclareSIUnit\ft{ft}
\DeclareSIUnit \ppb{ppb}
\DeclareSIUnit \ppt{ppt}
\DeclareSIUnit \samples{S}
\DeclareSIUnit \pe{PE}
\DeclareSIUnit \GeVmwe{GeV/mwe}
\DeclareSIUnit \mwe{mwe}



%% file: Commands.tex
\renewcommand{\arraystretch}{1.15}

\definecolor{lime}{HTML}{A6CE39}
\DeclareRobustCommand{\orcidicon}{
	\begin{tikzpicture}
	\draw[lime, fill=lime] (0,0) 
	circle [radius=0.16] 
	node[white] {{\fontfamily{qag}\selectfont \tiny ID}};
	\draw[white, fill=white] (-0.0665,0.095) 
	circle [radius=0.005];
	\end{tikzpicture}
	\hspace{-2mm}
}

\foreach \x in {A, ..., Z}{\expandafter\xdef\csname orcid\x\endcsname{\noexpand\href{https://orcid.org/\csname orcidauthor\x\endcsname}
			{\noexpand\orcidicon}}
   }
